\documentclass[prl,twocolumn,nofootinbib,showpacs]{revtex4}
\usepackage{bm}
\newcommand{\llangle}{\langle\!\langle}
\newcommand{\rrangle}{\rangle\!\rangle}
\newcommand{\LLangle}{\left\langle\!\!\!\left\langle}
\newcommand{\RRangle}{\right\rangle\!\!\!\right\rangle}
\begin{document}
\title{\bf Stochastic Schr\"odinger Equations with General Complex
Gaussian Noises}
\author{Angelo Bassi}
\email{bassi@ictp.trieste.it} \affiliation{The Abdus Salam
International Centre for Theoretical Physics \\ Strada Costiera,
11 --- 34014 Miramare--Grignano, Trieste, Italy, and \\
Istituto Nazionale di Fisica Nucleare, sezione di Trieste, Italy.}
\begin{abstract}
Within the framework of non Markovian stochastic Schr\"odinger
equations, we generalize the results of [Phys. Lett. A {\bf 224},
25 (1996)] to the case of general complex Gaussian noises; we
analyze the two important cases of purely real and purely
imaginary stochastic processes.
\end{abstract}
 \maketitle

In the last years, stochastic Schr\"odinger equations have
received considerable attention
\cite{pp,di,gpr,bel,bar,jp,ad1,str,Di2,ds,dgs,sdg,ydgs,bud,bg,gw}:
two are the basic reasons for this increasing popularity. On the
one hand, it has been shown (within the framework of {\it
dynamical reduction models}) \cite{pp,gpr,jp,ad1,bg} that by
adding commuting operators coupled to appropriate stochastic
noises into the Schr\"odinger equation, the statevector is driven
into one of the common eigenmanifolds of such
operators\footnote{With an appropriate choice of the operators and
noises, the reduction of the statevector occurs only at the
macroscopic level; the split between the micro and the macro is
defined by the parameters of the model.}; it is then possible to
combine the quantum evolution and the process of wavepacket
reduction into one single dynamical equation, providing a solution
to the well known measurement problem of quantum mechanics. On the
other hand, stochastic Schr\"odinger equations have proved to be a
novel and useful tool in the study of {\it open quantum systems},
i.e. quantum systems interacting with the surrounding environment
\cite{di,bel,bar,str,Di2,ds,dgs,sdg,ydgs,bud,pk,gw}. In this case,
there are two advantages with respect to the standard formalism
which resorts to the reduced density matrix of the open quantum
system (when all the degrees of freedom of the environment have
been traced away): from a theoretical point of view, it is
interesting to analyze how the statevector of such a system
evolves in time, following the different possible realizations of
the stochastic processes, while the reduced density matrix
formalism gives only an ensemble description of the system. From a
more ``practical'' point of view, in some cases it turns out to be
simpler to unfold the evolution of a stochastic Schr\"odinger
equation numerically on a computer and subsequently perform an
average over sufficiently many different trajectories, rather than
integrating numerically the equations for the corresponding
statistical operator. Moreover, the statistical operator
describing the ensemble of solutions of a stochastic Schr\"odinger
equation is, by construction, a positive definite operator; this
property is not always guaranteed by other approaches to the study
of open quantum systems.

In both cases, a stochastic Schr\"odinger equation takes the form
\begin{equation} \label{intw}
\frac{d}{dt}\, |\psi(t)\rangle \; = \; \left[ -\frac{i}{\hbar}\, H
+ f({\bf L}, {\bf z}(t)) \right] |\psi(t)\rangle,
\end{equation}
where $H$ is the free Hamiltonian of the system under study, while
$f({\bf L}, {\bf z}(t))$ is an appropriate function of the
operators ${\bf L} \equiv \{L_{1}, L_{2}, \ldots, L_{n} \}$ and of
the stochastic noises ${\bf z}(t) \equiv \{z_{1}(t), z_{2}(t),
\ldots, z_{n}(t) \}$.  The statistical operator is defined as the
ensemble mean, with respect to the noises ${\bf z}(t)$, of the
projection operators $|\psi(t)\rangle\langle\psi(t)|$:
\begin{equation}
\rho(t) \; = \; \llangle |\psi(t)\rangle\langle\psi(t)| \rrangle,
\end{equation}
where $\llangle{}\;\;\rrangle$ denotes the stochastic average. It
obeys the equation\footnote{Note that, in general, it is not
possible to write a closed equation for $\rho(t)$.}:
\begin{equation} \label{ints}
\frac{d}{dt}\, \rho(t) \; = \; -\frac{i}{\hbar} [H, \rho(t)] +
\llangle T[|\psi(t)\rangle\langle\psi(t)|] \rrangle,
\end{equation}
the form of the operator $T[\cdot]$ depending on the operators
${\bf L}$ and the statistical properties of the noises ${\bf
z}(t)$. In the case of statevector collapse models, $T[\cdot]$
describes the average effect of wavepacket reductions; in the case
of open quantum systems, it embodies the effect of the surrounding
environment on the system.

In the {\it white noise} case, equations (\ref{intw}) and
(\ref{ints}) take a simple form \cite{pp,di,gpr,bel,bar,jp,ad1},
and their most relevant properties have been extensively analyzed.
However, they do not encompass all physical interesting cases,
since they describe only Markovian evolutions.

Quite recently \cite{str,Di2,ds,dgs,sdg,ydgs,bud,bg,gw}, the
formalism of stochastic Schr\"odinger equations has been
generalized to the case of non--white complex Gaussian noises,
which describe non--Markovian evolutions. The basic equation is:
\begin{eqnarray} \label{gis1}
\frac{d}{dt}|\psi(t)\rangle & = & \left[ -\frac{i}{\hbar}\, H \; +
\; {\bf L}\cdot {\bf z}(t)\right.\nonumber \\ & & \left. - \; {\bf
L}^{\dagger} \cdot \int_{t_{0}}^{t} ds\, {\bf a}(t,s)\cdot
\frac{\delta}{\delta\, {\bf z}(s)} \right]|\psi(t)\rangle,
\end{eqnarray}
where ${\bf z}(t)$ are complex gaussian noises with zero mean and
correlation functions\footnote{The complex functions $a_{ij}(t,s)$
are the coefficients of the matrix ${\bf a}(t,s)$ appearing in
equation (\ref{gis1}).}:
\begin{eqnarray} \label{cf}
\llangle z^{\star}_{i}(t)\, z_{j}(s) \rrangle = a_{ij}(t,s) \qquad
\llangle z_{i}(t)\, z_{j}(s) \rrangle = 0 .
\end{eqnarray}
The dot $\cdot$ in equation (\ref{gis1}) denotes the product
between the corresponding tensorial quantities, e.g. ${\bf L}\cdot
{\bf z}(t) \equiv \sum_{i} L_{i}\, z_{i}(t)$.

Equation (\ref{gis1}) does not preserve the norm of the
statevector and, as such, it does not have a direct physical
meaning. It is possible to write \cite{dgs} an equation for the
normalized physical vectors $|\phi(t)\rangle = |\psi(t)\rangle /
\| |\psi(t)\rangle \|$; this equation is rather involved since it
does not have a closed form in $|\phi(t)\rangle$, and is non
linear. Anyway, for our purposes it suffices to work with
(\ref{gis1}), which gives the correct predictions once the
probability distribution $P[{\bf z}(t)]$ associated to the
stochastic process of (\ref{cf}) is replaced by:
\begin{equation} \label{ps}
Q[{\bf z}(t)] \; = \; P[{\bf z}(t)]\, \| |\psi(t)\rangle \|^{2}.
\end{equation}

Equation (\ref{gis1}) represents a breakthrough within the
stochastic Schr\"odinger equations formalism, since it can
describe a great variety of systems in many different physical
situations \cite{dgs,ydgs}. On the other hand, relations
(\ref{cf}) for the correlation functions set a limit for the
possible choices of the noise, as they put on the same ground both
the real and the imaginary parts of the complex noises\footnote{As
a matter of fact, the stochastic properties of ${\bf z}(t)$ are
invariant under phase transformation, i.e. ${\bf z}(t)$ and ${\bf
z}'(t) = e^{i\,\varphi} {\bf z}(t)$ share the same statistical
properties.} ${\bf z}(t)$: this is evident once we re--write the
correlation functions (\ref{cf}) in terms of ${\bf x}(t) =
\makebox{Re} [{\bf z}(t)]$ and ${\bf y}(t) = \makebox{Im} [{\bf
z}(t)]$:
\begin{eqnarray}
\llangle x_{i}(t)\, x_{j}(s) \rrangle \; = \; \llangle y_{i}(t)\,
y_{j}(s) \rrangle & = & \frac{1}{2}\, \makebox{Re}[a_{ij}(t,s)],
\nonumber \\
\llangle x_{i}(t)\, y_{j}(s) \rrangle \; = \; -\llangle y_{i}(t)\,
x_{j}(s) \rrangle & = & -\frac{1}{2}\, \makebox{Im}[a_{ij}(t,s)].
\nonumber
\end{eqnarray}
It is then interesting to analyze what happens in the case of more
general complex noises. There is a second reason why one would
take into account more general stochastic Schr\"odinger equations,
which is related to a characteristic feature of the formalism: in
general, there are infinitely many different stochastic
Schr\"odinger equations leading to the same equation for the
statistical operator and, as such, reproducing the same
statistical behavior. This is particularly simple to see in the
white noise case, which we now briefly consider; for simplicity,
we assume that the operators ${\bf L}$ are self--adjoint. In such
a case, equation (\ref{gis1}) becomes\footnote{Note that all white
noise equations are written as Stratonovich equations.} \cite{jp}:
\begin{equation} \label{gis2}
\frac{d}{dt}\,|\psi(t)\rangle \; = \; \left[ -\frac{i}{\hbar}\, H
\; + \; {\bf L}\cdot {\bf z}(t) - \frac{1}{2}\, {\bf L}^{2}
\right]|\psi(t)\rangle,
\end{equation}
where the correlation functions ${\bf a}(t,s)$ are now
$a_{ij}(t,s) = \delta_{ij}\, \delta(t-s)$. The above equation, as
previously mentioned, does not preserve the norm of the
statevector; for completeness, we write the corresponding
norm--preserving equation for the physical vectors
$|\phi(t)\rangle$:
\begin{eqnarray} \label{gis3}
\frac{d}{dt}|\phi(t)\rangle & = & \left[ -\frac{i}{\hbar}H + ({\bf
L} - \langle {\bf L} \rangle)\cdot ({\bf z}(t) + \langle {\bf
L}\rangle) \right. \nonumber \\
& & \left. -\frac{1}{2}( {\bf L}^{2} - \langle {\bf L}^{2}
\rangle) \right]|\phi(t)\rangle,
\end{eqnarray}
with $\langle {\bf L} \rangle = \langle \phi(t) | {\bf L} |
\phi(t) \rangle$. It is not difficult to prove that equation
(\ref{gis3}) or, equivalently, equation (\ref{gis2}) plus the
prescription (\ref{ps}), leads to the following equation for the
statistical operator\footnote{This is one of the basic equations
in the theory of decoherence \cite{jz,lib}; when the operators
${\bf L}$ commute among themselves, the non--Hamiltonian terms
determine the damping of the off--diagonal elements $\langle
\alpha | \rho(t) | \beta \rangle$, where $|\alpha\rangle,
|\beta\rangle$ are two different eigenstastes of ${\bf L}$.}:
\begin{equation} \label{jz}
\frac{d}{dt}\rho(t) = -\frac{i}{\hbar} \left[ H, \rho(t) \right] -
\frac{1}{2} \left[{\bf L} \left[{\bf L}, \rho(t)\right]\right],
\end{equation}
which of course is of the Lindblad type \cite{lind}.

We now show that equation (\ref{jz}) can be derived from {\it
infinitely} many different stochastic Sch\"odinger equations. As a
metter of fact, let us consider the noises ${\bf w}(t) = c\,{\bf
x}(t)$, where $c = e^{i\,\varphi}$ is a complex phase factor and
${\bf x}(t)$ is a {\it real} gaussian {\it white noise} with zero
mean and correlation function $\llangle x_{i}(t)\, x_{j}(s)
\rrangle \; = \; \delta_{ij} \, \delta(t-s)$, and the following
linear stochastic Schr\"odinger equation:
\begin{equation} \label{rex}
\frac{d}{dt} |\psi(t)\rangle \; = \; \left[ -\frac{i}{\hbar}\, H +
{\bf L}\cdot {\bf w}(t) -\, {\bf L}^{2} \cos^{2} \varphi
\right]|\psi(t)
\end{equation}
A little calculation shows that, for {\it any} value of $\varphi$,
equation (\ref{rex}) leads to (\ref{jz}). Note that, in order to
write equation (\ref{rex}), we have resorted to complex white
noises more general than those appearing in equation (\ref{gis2}),
since the correlation function $\llangle w_{i}(t)\, w_{j}(s)
\rrangle = c^{2}\,\delta_{ij}\, \delta(t-s)$ is no longer equal to
zero (compare with equation (\ref{cf})). Note also that, in spite
of the fact that they share the same statistical properties,
equations (\ref{gis2}) and (\ref{rex}) have a radically different
structure and lead to completely different time evolutions for the
statevector\footnote{In refs. \cite{gnd,gw} the same issue is
considered. In ref. \cite{gnd}, the authors prove that, in the
white noise case, the two stochastic Schr\"odinger equations of
the type (\ref{rex}) with $\varphi = 0$ (real noise) and $\varphi
= \pi/2$ (imaginary noise), lead to the same equation for the
statistical operator. In ref. \cite{gw} the authors analyze two
different unravelings for an open system interacting with a bath
of harmonic oscillators, corresponding to two different sorts of
measurements on the bath. The first unraveling leads to a
non--Markovian stochastic Schr\"odinger equation of the type
(\ref{gis1}), while the second unraveling leads to a
non--Markovian stochastic Schr\"odinger equation with a real
Gaussian noise (see equation (\ref{pp2r})). By construction, the
two equations describe (when the average is taken with respect to
the noise) the evolution of the same quantum open system.}.

This ``redundancy'' in the number of stochastic Schr\"odinger
equation leading to (\ref{jz}) has an important consequence: one
has the possibility to choose the most suitable equation according
to the specific problem under study. For example, if one wants to
reproduce the {\it statevector reduction process}, the best choice
is equation (\ref{gis2}), or equation (\ref{rex}) with $\varphi =
0$ (which corresponds to a purely real stochastic process):
\begin{equation} \label{ride}
\frac{d}{dt} |\psi(t)\rangle \; = \; \left[ -\frac{i}{\hbar} H \;
+ \; {\bf L} \cdot {\bf x}(t) \; - \; {\bf L}^{2} \right]
|\psi(t)\rangle;
\end{equation}
as already mentioned, the non--Hamiltonian terms of (\ref{ride}),
together with the prescription (\ref{ps}), drive the statevector
into one of the common eigenmanifolds of the operator {\bf L}, if
they commute among themselves.

If on the other hand one wants to take advantage of the
stochastic--Schr\"odinger--equation formalism to solve equation
(\ref{jz}), then the mathematically simplest equation is
(\ref{rex}) with $\varphi = \pi/2$ (i.e. a purely imaginary
noise):
\begin{equation} \label{wdr}
\frac{d}{dt} |\psi(t)\rangle \; = \; \left[ -\frac{i}{\hbar} H \;
+ \; i\, {\bf L} \cdot {\bf x}(t) \right] |\psi(t)\rangle.
\end{equation}
This equation is unitary, preserves the norm of the statevector
and thus coincides with the physical equation for the normalized
vectors $|\phi(t)\rangle = |\psi(t)\rangle/ \| |\psi(t)\rangle\|$.
The evolution it describes is different from that of equations
(\ref{ride}) and (\ref{gis2}), since it does not lead to the
localization of the statevector \cite{gnd}. Anyway, as long as one
is concerned only in time evolution of the density matrix,
equation (\ref{wdr}) is as good as (\ref{ride}) and (\ref{gis2})
(in particular, confront the physical, norm--preserving Eqs.
(\ref{wdr}) and (\ref{gis3})).

In the light of the above discussion, it becomes interesting to
generalize the stochastic Schr\"odinger equations formalism to
more general types of complex gaussian noises. To this purpose,
let us consider the following stochastic Schr\"odinger equation:
\begin{equation} \label{pp}
\frac{d}{dt}\,|\psi(t)\rangle \; = \; \left[ -\frac{i}{\hbar}\, H
\; + \; {\bf L}\cdot {\bf z}(t) + O \right]|\psi(t)\rangle,
\end{equation}
where, now, ${\bf z}(t)$ is a general complex gaussian stochastic
process with zero mean and correlation functions:
\begin{eqnarray}
\llangle z^{\star}_{i}(t)\, z_{j}(s) \rrangle  & = & a_{ij}(t,s)
\nonumber \\
\llangle z_{i}(t)\, z_{j}(s) \rrangle & = & b_{ij}(t,s);
\end{eqnarray}
$a_{ij}(t,s)$ and $b_{ij}(t,s)$ are generic functions, the only
requirement on them being the consistency conditions: $a_{ij}(t,s)
= a^{*}_{ji}(s,t)$ and $b_{ij}(t,s) = b_{ji}(s,t)$.

The operator $O$ is uniquely defined \cite{bud,bg} by the request
that (\ref{pp}) preservers the average value of the square norm of
$|\psi(t)\rangle$, this requirement being necessary in order to
work out a meaningful physical interpretation:
\begin{widetext}
\begin{eqnarray} \label{le}
\frac{d}{dt}\, \llangle \langle \psi(t)|\psi(t)\rangle\rrangle & =
& \llangle \langle\psi(t)| \left[ {\bf L}^{\dagger}\cdot {\bf
z}^{\star}(t) + O^{\dagger} + {\bf L}\cdot {\bf z}(t) + O \right]
|\psi(t)\rangle \rrangle \nonumber \\
& = & \llangle \langle\psi(t)| \,  O \, |\psi(t)\rangle \rrangle +
\LLangle \langle\psi(t)| \int_{t_{0}}^{t} ds\, \left( {\bf
L}^{\dagger}\cdot{\bf a}(t,s) + {\bf L}\cdot{\bf b}(t,s)
\right)\cdot \frac{\overrightarrow{\delta}}{\delta\,{\bf
z}(s)}|\psi(t)\rangle
\RRangle \; + \nonumber \\
& & \llangle \langle\psi(t)|\, O^{\dagger} \, |\psi(t)\rangle
\rrangle + \LLangle \langle\psi(t)| \int_{t_{0}}^{t} ds\,
\frac{\overleftarrow{\delta}}{\delta\,{\bf
z}^{\star}(s)}\cdot\left( {\bf a}^{\dagger}(t,s)\cdot{\bf L} +
{\bf b}^{\dagger}(t,s)\cdot{\bf L}^{\dagger} \right)
|\psi(t)\rangle \RRangle \; =  \; 0.
\end{eqnarray}
\end{widetext}
In the second line of equation (\ref{le}), the functional
derivative acts on the ket $|\psi(t)\rangle$, while in the third
line it acts only on the bra $\langle\psi(t)|$. In going from the
first to the second line, we have employed the Furutsu--Novikov
formula \cite{fn}. Equation (\ref{le}) is then satisfied if:
\begin{equation}
O \; = \; - \int_{t_{0}}^{t} ds\, \left({\bf L}^{\dagger}\cdot
{\bf a}(t,s) + {\bf L}\cdot{\bf b}(t,s) \right)\cdot
\frac{\delta}{\delta\,{\bf z}(s)},
\end{equation}
and equation (\ref{pp}) becomes:
\begin{eqnarray} \label{pp2}
\lefteqn{\frac{d}{dt}\,|\psi(t)\rangle \; = \; \left[
-\frac{i}{\hbar}\, H \;
+ \; {\bf L}\cdot {\bf z}(t) \right.} \nonumber \\
& & \left. - \int_{t_{0}}^{t} ds\, \left( {\bf L}^{\dagger}\cdot
{\bf a}(t,s) + {\bf L}\cdot{\bf b}(t,s) \right)\cdot
\frac{\delta}{\delta\,{\bf z}(s)}\right]|\psi(t)\rangle.\qquad
\end{eqnarray}
We can easily see that equation (\ref{pp2}) generalizes
(\ref{gis1}) and correctly reduces to it when ${\bf b}(t,s) = {\bf
0}$. One can write the equation for the evolution of the
statistical operator and work out a perturbation expansion in more
or less the same way it has been done for the ${\bf b}(t,s) = {\bf
0}$ case \cite{ydgs,gw}, but this goes beyond the scope of the
present article.

It is instructive to analyze in greater detail the important case
of self--adjoint operators ${\bf L}$:
\begin{eqnarray} \label{pp3}
\lefteqn{\frac{d}{dt}\,|\psi(t)\rangle \; = \; \left[
-\frac{i}{\hbar}\, H \;
+ \; {\bf L}\cdot {\bf z}(t) \right.} \nonumber \\
& & \left. - {\bf L}\cdot\int_{t_{0}}^{t} ds\, \left({\bf a}(t,s)
+ {\bf b}(t,s) \right)\cdot \frac{\delta}{\delta\,{\bf
z}(s)}\right]|\psi(t)\rangle,\qquad
\end{eqnarray}
together with the limit ${\bf x}(t) \rightarrow {\bf 0}$ and ${\bf
y}(t) \rightarrow {\bf 0}$; note that one cannot take such limits
for equation (\ref{gis1}), due to the particular form of the
correlation functions (\ref{cf}). In the first case (a purely
imaginary noise), ${\bf a}(t,s) = - {\bf b}(t,s)$, so that the
third term at the right hand side disappears and (\ref{pp3})
coincides with (\ref{wdr}). This is an expected result, since a
unitary evolution preserves the norm of the statevector and thus
it does not need any corrective term. In the second case (a purely
real noise), ${\bf a}(t,s) = {\bf b}(t,s)$ and (\ref{pp3})
becomes:
\begin{eqnarray} \label{pp2r}
\frac{d}{dt}\,|\psi(t)\rangle & = & \left[ -\frac{i}{\hbar}\, H \;
+ \; {\bf L}\cdot {\bf z}(t) \right. \nonumber \\
& & \left. - 2{\bf L}\cdot\int_{t_{0}}^{t} ds\, {\bf a}(t,s)\cdot
\frac{\delta}{\delta\,{\bf z}(s)}\right]|\psi(t)\rangle,\qquad
\end{eqnarray}
This equation has been studied in detail in ref. \cite{bg}, where
it has been proved that --- as for the white noise equation
(\ref{ride}) --- the non Hamiltonian terms drive the statevector
into one of the common eigenmanifolds of the operators ${\bf L}$,
when they commute among themselves.

To summarize, we have generalized the formalism of non--Markovian
stochastic Schr\"odinger equations to any kind of complex Gaussian
noises, recovering the results of
\cite{str,Di2,ds,dgs,sdg,ydgs,bud,bg} as a special case. We have
seen that such a formulation is more ``flexible'' as it covers the
two particular but very important cases of purely real and purely
imaginary noises, and in general it allows to take the most
suitable form of the noise according to the specific problem under
study.


\begin{thebibliography}{99}
\bibitem{gpr} G. C. Ghirardi, P. Pearle and A. Rimini, Phys. Rev. A
{\bf 42}, 78 (1990). G.C. Ghirardi, R. Grassi and P. Pearle,
Found. Phys. {\bf 20}, 1271 (1990). G.C. Ghirardi, R. Grassi and
A. Rimini, Phys. Rev. A {\bf 42}, 1057 (1990). G.C. Ghirardi, R.
Grassi and F. Benatti, Found. Phys. {\bf 25}, 5 (1995).
\bibitem{pp} P. Pearle, Phys. Rev. Lett. {\bf 53}, 1775 (1984).
Phys. Rev. A {\bf 39}, 2277 (1989).
\bibitem{di} L. Di\'osi, Phys. Lett. A {\bf 132}, 233
(1988). J. Phys. A {\bf 21}, 2885 (1988). Phys. Rev. A {\bf 42},
5086 (1990).
\bibitem{bel} V.P. Belavkin, in {\it Lecture Notes in Control and
Information Science} {\bf 121}, A. Blaqui\`ere ed., 245 (1988).
V.P. Belavkin and P. Staszewski, Phys. Rev. A {\bf 45}, 1347
(1992).
\bibitem{bar} A. Barchielli, Quantum Opt. {\bf 2}, 423
(1990). Int. J. Theor. Phys. {\bf 32}, 2221 (1993).
\bibitem{jp} N. Gisin and J. Percival, Phys. Lett. A {\bf
167}, 315 (1992). J. Phys. A {\bf 25}, 5677 (1992). J. Phys. A
{\bf 26}, 2233 (1993); {\bf 26}, 2243 (1993). I. Percival, {\it
Quantum State Diffusion}, Cambridge University Press, Cambridge
(1998).
\bibitem{ad1} L.P. Hughston, Proc. R. Soc. London A {\bf 452},
953 (1996). S.L. Adler and L.P. Howritz: Journ. Math. Phys. {\bf
41}, 2485 (2000). S.L. Adler and T.A. Brun: Journ. Phys. A {\bf
34}, 4797 (2001). S.L. Adler, Journ. Phys. A {\bf 35}, 841 (2002).
D.C. Brody and L.P. Hughston, Proc. R. Soc. London A {\bf 458}
(2002).
\bibitem{str} W.T. Strunz, Phys. Lett. A {\bf 224}, 25
(1996).
\bibitem{Di2} L. Di\'osi, Quantum Semiclass. Opt. {\bf 8},
309 (1996).
\bibitem{ds} L. Di\'osi and W.T. Strunz, Phys. Lett. A {\bf
2358}, 569 (1997).
\bibitem{dgs} L. Di\'osi, N. Gisin and W.T. Strunz, Phys.
Rev. A {\bf 58}, 1699 (1998).
\bibitem{sdg} W.T. Strunz, L. Di\'osi and N. Gisin, Phys.
Rev. Lett. {\bf 82}, 1801 (1999). L. Di\'osi, N. Gisin, W.T.
Strunz and T. Yu, Phys. Rev. Lett. {\bf 83}, 4909 (1999).
\bibitem{ydgs} T. Yu, L. Di\'osi, N. Gisin and W.T. Strunz,
Phys. Rev. A {\bf 60}, 91 (1999). Phys. Lett. A {\bf 265}, 331
(2000).
\bibitem{bud} A.A. Budini, Phys. Rev. A {\bf 63}, 012106
(2000).
\bibitem{bg} A. Bassi and G.C. Ghirardi, Phys. Rev. A {\bf 65},
042114 (2002).
\bibitem{gw} J. Gambetta and H.M. Wiseman, Phys. Rev. A {\bf 66},
012108 (2002). J. Gambetta and H.M. Wiseman, to appear in Phys.
Rev. A (preprint {\it quant--ph/0208169}).
\bibitem{pk} M.B. Plenio and P.L. Knight, Rev. Mod. Phys.
{\bf 70}, 101 (1998).
\bibitem{jz} E. Joos and H.D. Zeh, Zeit f\"ur Phys. B {\bf 59},
233 (1985).
\bibitem{lib} D. Giulini, E. Joos, C. Kiefer, J. Kupsch, I.--O.
Stamatescu and H.D. Zeh, {\it Decoherence and the Appearance of a
Classical World in Quantum Theory}, Springer (1996).
\bibitem{lind} G. Lindblad, Commun. Math. Phys. {\bf 48}, 119
(1976).
\bibitem{gnd} G.C. Ghirardi and R. Grassi, in: {\it  Nuovi
problemi della logica e della filosofia della scienza}, CLUEB,
Bologna (1991).
\bibitem{fn} A. Novikov, Zh. Eksp. Teor. Fiz. {\bf 47}, 1915
(1964) [Sov. Phys. JETP {\bf 20}, 1290 (1965)]. K. Sobczyk, {\it
Stochastic Wave Propagation}, Elsevier, Amsterdam {1985}.
\end{thebibliography}
\end{document}